\documentclass[a4paper]{jpconf}

\usepackage{graphicx,iopams}

\newcommand{\PP}{{\it Preprint}\ }

\newcommand{\be}{\begin{equation}}
\newcommand{\ee}{\end{equation}}
\newcommand{\reff}[1]{(\ref{#1})}
\newcommand{\kb}{\ensuremath k_{\rm B}}

\newcommand{\EE}{{\mathcal E}}
\newcommand{\FF}{{\mathcal F}}
\newcommand{\op}{\phi}

\begin{document}

\title{The Casimir effect: from quantum to critical fluctuations}

\author{Andrea Gambassi}

\address{Max-Planck-Institut f\"ur Metallforschung,
Heisenbergstr. 3, D-70569 Stuttgart, Germany
and
Institut f\"ur Theoretische und Angewandte Physik,
Universit\"at Stuttgart,
Pfaffenwaldring 57, D-70569 Stuttgart, Germany.}

\ead{gambassi@mf.mpg.de}

\begin{abstract}
The Casimir effect in quantum electrodynamics (QED) is perhaps the
best-known example of fluctuation-induced long-ranged force acting on
objects (conducting plates) immersed in a fluctuating medium (quantum
electromagnetic field in vacuum).
A similar effect emerges in statistical physics, where the force
acting, e.g., on colloidal particles immersed in a binary liquid mixture is
affected by the classical thermal fluctuations occurring in
the surrounding medium. The resulting Casimir-like force acquires universal
features upon approaching a critical point of the medium and becomes
long-ranged at criticality. In turn, this universality allows one to
investigate theoretically the temperature dependence of the force via
representative models and to stringently test the corresponding
predictions in experiments.
In contrast to QED, the Casimir force resulting from critical fluctuations
can be easily tuned with respect to strength and sign by surface
treatments and temperature control.
We present some recent advances in the theoretical study of the universal
properties of the critical Casimir force arising in thin films. The
corresponding  
predictions compare very well with the experimental results obtained for
wetting layers of various fluids. We discuss how the
Casimir force between a colloidal particle and a planar wall immersed in a
binary liquid mixture has been measured with femto-Newton accuracy,
comparing these experimental results with the corresponding theoretical
predictions.
\end{abstract}

\section{Introduction}

This contribution focuses on the Casimir effect which arises when one confines
fluctuations of a different nature compared to those
originally considered by Casimir in his pioneering work~\cite{Casimir}. 
Indeed we shall be concerned here with the fluctuations of thermal origin
which  occur close to a second-order phase transition point ({\it critical
  point}). The resulting Casimir effect is apparently less widely
known than the corresponding one within quantum electrodynamics (QED). 
Nonetheless, the resulting femto-Newton forces at the sub-micrometer scale can
be put to work in soft matter systems and their high degree of tunability
might even be exploited for concrete applications in colloidal suspensions. 
However, before focussing on the Casimir effect due to critical fluctuations 
we first discuss in subsec.~\ref{subsec:QED} the effect within QED, 
following in spirit the original derivation by
Casimir~\cite{Casimir}. In subsec.~\ref{subsec:ccf} we then explain how this
effect arises in statistical physics upon approaching a critical point,
highlighting the analogies and differences between the two.
In sec.~\ref{sec:evid} we present an overview of the available theoretical
predictions which are relevant for the qualitative and quantitative 
interpretation of the experimental results. A short summary,
with perspectives and applications is presented in sec.~\ref{sec:persp}.

\section{The Casimir effect }

\subsection{The effect in Quantum Electrodynamics}
\label{subsec:QED}
The Casimir effect is named after Hendrik Casimir who discovered, 60 years ago,
that -- quite surprisingly -- two parallel, perfectly conducting and uncharged
metallic plates in vacuum  {\it attract} each other due to the 
{\it quantum fluctuations} of the electromagnetic fields, even at zero
temperature $T=0$~\cite{Casimir}.
Indeed the plates (assumed to be perfectly conducting)
effectively impose boundary conditions (BCs) on the
electromagnetic fields so that, more specifically, 
${\bf E}_\|$, ${\bf B}_\bot = 0$ where 
${\bf E}_\|$ and ${\bf B}_\bot$ are the  components of the electric ${\bf E}$
and magnetic ${\bf B}$ fields which are parallel and transverse to the surface
of the plates. 
As a result of these boundary conditions, the {\it fluctuation modes} of the
fields in the space within the two plates can only have a specific
set of $L$-dependent allowed wave-vectors, where $L$ is the separation between
the parallel plates. 
For example, in the case of Dirichlet BCs, the component
$k_\bot$ of the wave-vector ${\bf k}$ 
perpendicular to the plates assumes only the quantized values 
$k_\bot = \pi n/L$ 
with $n=1,2,\ldots$.
Roughly speaking, the ``unbalance'' between the pressure exerted by the
allowed modes within the plates and the one exerted by the modes outside them
is at the origin of the Casimir effect.
This statement can be made quantitative by calculating the
size-dependent energy 
$\EE(L)$ of the electromagnetic fields allowed in the portion of the vacuum
within the plates, which is given, according to QED, by
\be
\EE(L) = \sum_{\rm modes} \frac{1}{2} \hbar c |{\bf k}_{\rm modes}|\,,
\ee
where ${\bf k}_{\rm modes}$ are the corresponding 
wave-vectors and $c$ is the speed of light.
%
%
For perfectly conducting plates of large transverse area $S$, the expansion of
$\EE(L)$ in decreasing powers of $L$ takes the form (in three spatial
dimensions)
\be
\EE(L) = \EE_{\rm bulk} + \EE_{\rm surf}^{(1+2)} + S \frac{\hbar c}{L^3}
\left[ -\frac{\pi^2}{1440} + O((\kappa L)^{-2})\right]\,,
\label{eq:eqed}
\ee
where $\EE_{\rm bulk} \propto S L^1$ is the energy associated to the
electromagnetic field of the vacuum in the absence of the plates within the
volume of space $V=SL$ enclosed by them; $\EE_{\rm surf}^{(1+2)} = \EE_{\rm
  surf}^{(1)}  + \EE_{\rm
  surf}^{(2)} \propto S L^0$ is the sum of the energies $\EE_{\rm surf}$
associated to the introduction of each single plate, {\it separately}, in the
vacuum; The next term in the expansion~\reff{eq:eqed}, i.e.,
\be
\EE_{\rm Cas}(L)  \equiv  -\frac{\pi^2}{1440}\, S \frac{\hbar c}{L^3}\,,
\label{eq:ecas}
\ee 
is proportional to $SL^{-3}$ and represents the {\it
  interaction energy} between the two plates which is due to their {\it
  simultaneous} presence in space. This is the 
Casimir term we are interested
in and which is responsible for the celebrated Casimir force.
Further terms in the expansion are of order $O((\kappa L)^{-2})$ compared
to $\EE_{\rm Cas}$, where $\kappa$ is a material-dependent parameter which
describes the deviations of the plates from the perfectly
conducting behaviour due to the {\it finite} conductivity of the metal.
In the expansion~\reff{eq:eqed} the first term $\EE_{\rm bulk}$ is
a property of the vacuum, the second one ($\EE_{\rm surf}^{(1+2)}$) depends on
the material properties of the plates, whereas the third
one, $\EE_{\rm Cas}$ in eq.~\reff{eq:ecas}, 
is {\it universal} in that it does not depend on the
specific material which the plates are made of, but only on geometrical
properties ($S$ and $L$) and on 
fundamental constants ($\hbar c$).  The relevance
of higher-order terms in the expansion~\reff{eq:eqed} depends on the material
parameter $\kappa$.  

The Casimir term, the third of the 
the expansion of the vacuum
energy $\EE(L)$ in decreasing powers of $L$, actually gives rise to a
measurable effect: a small displacement $\delta L$ of one of the two plates
results in a change $\delta \EE(L)$ in the energy
of the fields within the plates and therefore in 
a force $F_{\rm in}(L) = -\delta\EE/\delta L$ (pushing on the inner
surface of each single plate).
Focussing on one of the two plates, the contribution  to the force $F_{\rm
  in}(L)$ due to the change of $\EE_{\rm bulk}$ in $\EE(L)$, i.e., 
$F_{\rm bulk} \equiv
-\delta \EE_{\rm bulk}/\delta L$, is actually canceled out by the force $F_{\rm
  out} = F_{\rm in}(L=\infty) = F_{\rm  bulk}$ acting from the other side of
the plate and due to the fluctuation modes in the corresponding 
half-space outside the plate. 
$\EE_{\rm  surf}$ does not change upon changing $L$ so that the total force
$F(L) = F_{\rm in}(L) - F_{\rm out}$ acting on each single plate is
due only to the change in the Casimir energy:
\be
\frac{F(L)}{S} = 
-\frac{\pi^2}{480} \frac{\hbar c}{L^4} \quad\mbox{for}\quad L \gg
  \kappa^{-1}\,.
\label{eq:fcas}
\ee
The universal behaviour encoded in this equation can only be detected for a
separation $L$ between the plates which is much larger than the length scale
$\kappa^{-1}$, being material-specific
properties relevant at smaller distances. 
In particular, $\kappa$ is related to the
typical wave-vector at which the plates are no longer effective
in imposing the boundary conditions on the fields. As a rough
estimate, this scale is set by the plasma frequency $\omega_p$ of the metal
(more detailed calculations support this 
argument~\cite{exp-cas}), 
i.e., $c \kappa \simeq \omega_p$, where $\omega_p \simeq
3\cdot 10^{15}$Hz for copper, yielding $\kappa^{-1} \simeq 0.3\mu$m.
Attempts to verify experimentally eq.~\reff{eq:fcas} started already in
1958~\cite{Spar-58} but the first sound experimental confirmation came only 40
years later within the range of separations $0.5\mu {\rm m} \lesssim L
\lesssim 6\mu{\rm m}$~\cite{exp-cas,exp-PD}. 
Apart from the relevance of this effect for possible technological
applications in micro- and nano-electromechanical systems (MEMS and NEMS,
respectively), in which the associated force is responsible for 
{\it stiction}, its precise
measurement could be useful in order to 
test of the validity of some fundamental
laws down to the nano-meter scale, see, e.g., ref.~\cite{Cas-rev}.  

\subsection{The effect in Statistical Physics}
\label{subsec:ccf}
\begin{figure}[h!!]
\begin{center}
\includegraphics[scale=1.3]{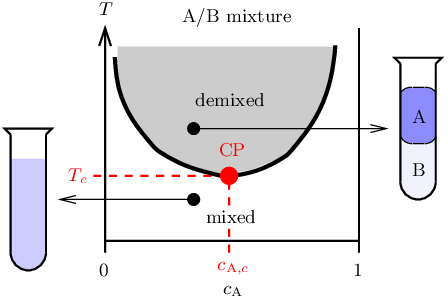}
\end{center}
\caption{\label{fig:PD}%
Schematic phase diagram of a binary liquid mixture with a lower critical point
(CP). 
On the left and on the
right of the diagram, the schematic side views of a test tube filled with the
binary liquid mixture respectively in the mixed and in the demixed phase are
shown.
}
\end{figure}

Thirty years after the seminal paper by Casimir, M.~E.~Fisher and
P.~-G.~de Gennes published a note "On the phenomena at the walls in a critical
binary mixture"~\cite{FdG} in which it was shown that Casimir-like effects
(i.e., {\it fluctuation-induced forces}) arise also in statistical
physics when a medium in which fluctuations of a certain nature take place is
spatially confined~\cite{KG-99}. 
In order to understand how these effects arise and what their relation is with
the Casimir effect described in subsec.~\ref{subsec:QED}, 
we briefly recall here some basics facts about binary
mixtures, their critical points and the effect of confinement. 
The schematic phase diagram of a liquid binary mixture in the bulk 
is depicted in
fig.~\ref{fig:PD}. For a  fixed pressure, 
the relevant thermodynamic variables are the temperature $T$ and
the mass fraction $c_A$ of one of the two components in the mixture. 
%
%
%
%
%
%
%
For a certain concentration $c_A$, the two components of the
mixture are mixed at low temperatures and 
the resulting liquid solution looks homogeneous in a test
tube (schematically represented on the left of the phase diagram). Upon
increasing the temperature, however, the liquid demixes into an A-
and a B-rich solution and becomes inhomogeneous in the test tube (represented
on the right), where the two solutions are typically separated by an
interface. The transition between the mixed and the demixed phases
occurs at the solid  first-order transition line. At the lowest point (CP) of
this line the transition becomes critical, i.e., second-order. This point is
referred to as the (lower) {\it critical point}.
The mixed and demixed phases can be distinguished by monitoring the so-called
{\it order parameter} of the transition, which can be identified with the
deviation $\delta c_A({\bf x}) \equiv c_A({\bf x}) - c_A$ of the {\it local}
concentration $c_A({\bf x})$ of, say, component A from its average value in
the mixture. Indeed, the thermal average $\langle \delta c_A({\bf x})\rangle$ 
of the order parameter is homogeneous in the mixed phase, whereas it assumes
two different values in the A-rich and B-rich solution which coexist in the
demixed phase. The order parameter is characterized by a spatial correlation
length $\xi$, so that, within each single phase,
$\langle\delta c_A({\bf x})
\delta c_A({\bf x}') \rangle -
\langle\delta c_A({\bf x}) \rangle  \langle\delta c_A({\bf x}')\rangle \propto
\exp\{-|{\bf x} - {\bf x}'|/\xi\}$, where $\xi$ depends on $T$ and $c_A$. Upon
crossing the first-order transition line, the correlation length stays
finite, whereas it diverges upon approaching the critical point. 
The physical behaviour of the system at the scale set by $\xi$ is actually
determined by the fluctuations of the order parameter, which becomes the
relevant physical quantity for the description of the system. In terms of
it, one can express the effective free energy $\FF$ of the
system and determine the behaviour of the thermodynamic properties upon
approaching the transition. 

In order to understand the consequences of confining such a binary liquid
mixture, we have first to consider the result of inserting 
a single plate (made up of, say, glass) into the mixture.
In this case, due to expected quantitative differences in the interactions
between the molecules of the plate substrate 
and those of the two components of the
mixture, the plate shows preferential adsorption for one of the two
components, say, A. Accordingly, the presence of the plate in the mixture
induces a local increase of the average order parameter $\langle \delta
c_A({\bf x})\rangle$ close to it. Upon increasing the correlation length
$\xi$ (i.e., getting closer to the critical point), the plate effectively
imposes boundary conditions on the order parameter~\cite{bind:83,diehl:86:0}.

\begin{table}[h!!]
\caption{\label{tab:comp}Analogies and differences between the Casimir effect
  in (zero-temperature) 
  QED and in Statistical Physics (see also footnote~\ref{footnote1}).}
\begin{center}
\begin{tabular}{rc|c}
\br
 & QED & Stat. Phys\\
\mr
& \hspace{3.5cm} & \hspace{3.5cm} \\[-4mm]
fluctuating quantity: & ${\bf E}$, ${\bf B}$ & order parameter $\op$\\[2mm]
excitation: & Quantum & Thermal(classical) \\
& $\hbar c$ ($T=0$) & $\kb T$ ($\hbar =0$) \\[2mm]
range of fluct.: & $\infty$ & finite: $\xi$ \\
&& $\xi\nearrow \infty$ close to CP \\
& $\Downarrow$ & $\Downarrow$ \\
& \multicolumn{2}{c}{{\bf \fbox{\hspace{1.8cm}Confinement\hspace{1.8cm}}}} \\
& $\Downarrow$ & $\Downarrow$ \\
& long-ranged force & range: $\xi$ \\
&& long-ranged at CP \\
\br
\end{tabular}
\end{center}
\end{table}

Now that we have recalled how to describe the binary mixture in terms of
a fluctuating order parameter (generically refereed to as $\op$ in what
follows) and the effect of plates in terms of boundary conditions, it is easy
to understand in which sense a Casimir-like interaction arises in these
instances.
Indeed (see table~\ref{tab:comp}), 
the Casimir effect in QED emerges because there are some 
quantities (the fields ${\bf E}$ and ${\bf B}$) which ``fluctuate'' in space
and time due to the quantum nature of the ``medium'' (actually the vacuum) in
which their fluctuations take place. 
Accordingly, the relevant scale of the phenomenon is
set by $\hbar c$ at zero temperature. 
The two-point correlation function of the fluctuations of these quantities are
characterized by an algebraic decay in space, i.e.,
the associated range is infinite (due to the vanishing mass $m_\gamma=0$ of
the photon). When the spectrum of these fluctuations gets modified by external
bodies (metallic plates) imposing boundary conditions on the fluctuating
fields, the associated energy $\EE$ turns into a function of 
the position of these
boundaries and as a result an effective long-ranged Casimir force arises on
them.
In statistical physics, close to a critical point, the relevant fluctuating
quantity is the order parameter $\op$
of the phase transition ($\op({\bf x}) = \delta c_A({\bf x})$ in
the case of the binary mixture) and its
fluctuations are of thermal nature and due to the coupling to the thermal bath
at temperature $T$. Accordingly, the relevant scale of the phenomenon is $\kb
T$ and the fluctuations are of classical nature (i.e., $\hbar$ plays no
role). The range of these fluctuations is  given by $\xi$ and therefore is {\it
  finite} but tunable upon approaching the critical point. As in the case of
QED, external bodies impose boundary conditions on the order parameter $\op$
and therefore they affect the spectrum of its allowed fluctuations and the
associated effective free energy $\FF$ depends on the positions of these
bodies. As a result, a force is expected to act on them, with a range set by
the correlation length $\xi$. 
In table~\ref{tab:comp} we summarize analogies and differences 
between the Casimir
effect occurring in QED and in statistical
physics\footnote{\label{footnote1}In some cases, the correspondence
  illustrated in table~\protect{\ref{tab:comp}} goes beyond the mere
  analogy. Indeed, for the simplest geometrical settings, such as film
  geometry, there is a mapping
  between the Casimir effect in QED in $d$ spatial dimensions and the effect
  at the critical point of the so-called Gaussian model in $d+1$ dimensions.}.
Even though we have been referring to the case of a binary liquid mixture, the
line of argument presented above extends to critical phenomena in general. 
Indeed, consider a medium close to its critical point and confined between two
parallel plates at a distance $L$ (film geometry). The medium might be, e.g., 
pure $^4$He or a $^3$He-$^4$He mixture close to the superfluid transition, a
binary liquid mixture, a Bose gas close to the Bose-Einstein condensation,
liquid crystals etc. 
If the correlation length $\xi$ of the fluctuations of the order parameter 
and the thickness $L$ of
the film are much larger than the microscopic length scale $\ell_{\rm micr}$
(set, say, by the molecular scale of the system), the {\it free energy}
$\FF$ of the confined medium can be decomposed in decreasing powers of $L$ for
a fixed value of $L/\xi$ and large transverse area $S$, in analogy to
eq.~\reff{eq:eqed}~\cite{krech:99:0,D-book} 
\be
\FF(T,L) = \FF_{\rm bulk} + \FF_{\rm surf}^{(1+2)} + S \frac{\kb T}{L^2}
\Theta_\|(L/\xi) + {\rm corr.}
\label{eq:fsp}
\ee 
We have assumed here that the only relevant thermodynamic variable is the
temperature $T$, which determines the correlation length 
$\xi \sim \xi_0|(T-T_c)/T_c|^{-\nu}$ where $\xi_0$ is a system-specific
quantity and $\nu$ a {\it universal} exponent, in the sense specified below. 
Possible additional thermodynamic variables, such as the concentration
$c_A$ of the binary mixture, are assumed to be tuned to their critical values,
i.e., $c_A = c_{A,c}$ with reference to fig.~\ref{fig:PD}.
In the expansion~\reff{eq:fsp} the first term $\FF_{\rm bulk}$ is the free
energy of the {\it bulk} medium in a volume $S\times L$, i.e., 
$\FF_{\rm bulk} \propto S L$,
the second one ($\FF_{\rm surf}^{(1+2)}$) represents the sums of the 
free energy costs $\FF_{\rm surf}^{(i)} \propto S$ 
for the separate introduction of the two
walls in the system, i.e.,  $\FF_{\rm surf}^{(1+2)} = \FF_{\rm surf}^{(1)} +
\FF_{\rm surf}^{(2)}$. The third term represents, as in eq.~\reff{eq:eqed}, 
the {\it interaction (free)energy} between the two walls. Compared to the
analogous term in the case of QED we note, as expected, that $\kb T$ replaces
$\hbar c$ in providing the scale of the phenomenon~\footnote{Due to the
  different engineering 
  dimensions of $[\hbar c] = {\rm energy}\times{\rm length}$ and
  $[\kb T] = {\rm energy}$, and taking into account the expected
  proportionality to $S$, the two effects are characterized by different
  powers in their dependence on $L$.} %
and that the interaction between the two walls is no longer long-ranged but
its {\it range is set by $\xi$}  via the dependence on $L/\xi$ of the scaling
function $\Theta_\|$(\footnote{The scaling function $\Theta_\|$ actually
  depends also on whether a certain correlation length $\xi$ is realized above
or below the bulk critical temperature $T_c$. This is made clear later on by
expressing the scaling function as a function of an appropriate scaling
variable $x$,
c.f., eq.~\reff{eq:defx}, instead of $u=L/\xi$, as done here.}).
This dependence can be understood as follows: the perturbation that the wall
induces on the order parameter extends within a distance $\sim \xi$ far from
it, so that the two walls can interact with each other only at separation
$L\lesssim \xi$ whereas
the interaction energy vanishes for $L \gg \xi$,
so that $\Theta_\|(x\rightarrow \infty) = 0$.

The scaling function $\Theta_\|$ is {\it universal} (as the analogous term
$\EE_{\rm Cas}$~\reff{eq:ecas} in eq.~\reff{eq:eqed}) in the
sense that it depends only on some gross features of: 
\begin{itemize}
\item[(a)] the system in the
bulk, such as the range and symmetries of the interaction and the kind of order
parameter which describes the phase transitions,
\item[(b)] the surfaces which provide the confinement, such as the
possible symmetries of the bulk system which they break, e.g., by favoring
certain values of the order parameter at the boundaries. These preferences
eventually translate into effective boundary conditions for the order
parameter~\cite{bind:83,diehl:86:0}. In addition, $\Theta_\|$ depends on
the shape of the boundaries, as we shall see below. 
\end{itemize}
The general features at points (a) and (b) define the so-called {\it bulk} and
{\it surface universality classes} of the confined system, respectively. 
This universality is typical of critical phenomena and allows one to
investigate universal properties such as $\nu$ and $\Theta_\|$ by means of
suitable representative models 
belonging to the same bulk and
surface universality classes of the actual system one is interested in and
which lend themselves for a simpler theoretical analysis. 
Analogously to what happens in QED, a small change $\delta L$ of the distance
between the confining surfaces leads to a change $\delta \FF$ in the free
energy and therefore to a
force $-\delta \FF/\delta L$ acting on the displaced wall. The $L$-independent
contribution to this force due to $\FF_{\rm bulk}$ in eq.~\reff{eq:fsp} is
counterbalanced by the same contribution acting from outside the walls when
they are immersed in the critical medium, so that the net force $F(L,T)$ is
given by (see eq.~\reff{eq:fsp})
\be
\frac{F(T,L)}{S} = \frac{\kb T}{L^3} \vartheta_\|(L/\xi) \quad\mbox{for}\quad
L, \, \xi \gg \ell_{\rm micr},
\label{eq:critfcas}
\ee
where $\vartheta_\|(u) = -2 \Theta_\|(u) + u \Theta_\|'(u)$ and it shares the
properties of universality with $\Theta(u)$. The universal
scaling function $\vartheta_\|$ can be conveniently determined from the
analysis of suitable representative models, either via field-theoretical
methods or Monte Carlo simulations. In turn, the resulting predictions can be
very stringently compared with the experimental determination of
$\vartheta_\|$.


\subsection{The force of quantum and critical fluctuations}

In the preceding two sections we have briefly reviewed the origin of the
Casimir effects due to quantum and critical fluctuations, highlighting the
analogy and differences between them and the corresponding scaling and
universal properties. In view of possible
experimental investigations, however, 
it is important to estimate the expected
magnitude of these effects. 
Assuming (quite unrealistically, indeed) that one is able to realize
a system of two parallel plates of area $S = 1\,{\rm cm}^2$ at a distance $L =
1\, \mu$m, the zero-temperature Casimir force resulting from quantum
fluctuations is roughly estimated as 
$\sim S \times \hbar c/L^4 \simeq 6 \cdot 10^{-7}\,$N, 
whereas the force due to critical fluctuations
possibly occurring at room temperature ($T\simeq 300\,$K) can be estimated as 
$\sim S \times \kb T/L^3 \simeq 4 \cdot 10^{-7}\,$N. Even though these
appear to be
quite small forces, their magnitudes are actually comparable with the weight
$\simeq 2\cdot 10^{-7}\,$N of a water droplet of half a millimeter in
diameter. 
However, the realization of such a geometrical setting is actually extremely
difficult due to the problem 
of maintaining the alignment between the two plates
within the required accuracy. 
This problem is particularly difficult when trying
to detect the Casimir effect in quantum electrodynamics and indeed  this has
been achieved in the
parallel-plate geometry only quite recently~\cite{exp-PD}.  
%
%
%
For the detection of the critical Casimir effect
the problem of alignment can be solved by taking advantage of physical
phenomena, such as {\it wetting}, which naturally lead to the formation of
confined films of fluids of a certain controllable well-defined thickness $L$, 
which can be varied by acting on the thermodynamic parameters of the system, 
such as pressure and temperature~\cite{dietrich,krech:92wet}.

\subsection{Wetting films and critical endpoints}
\label{subsec:cep}

\begin{figure}
\begin{center}
\includegraphics[scale=1.15]{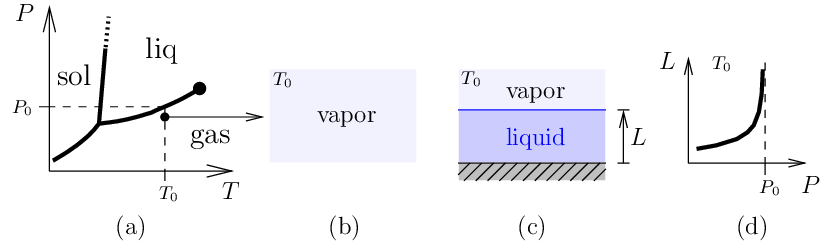}
\end{center}
\caption{\label{fig:wetfilm}%
Formation of a  {\it complete wetting} film of a fluid. 
Close to the liquid-gas phase
transition (a), a vapour (b) condenses in the presence of a
suitably chosen (solid) substrate, forming a liquid film (c), the thickness
$L$ of which diverges upon approaching the bulk condensation pressure $P_0$
(d). 
}
\end{figure}

In fig.~\ref{fig:wetfilm}(a) we depict schematically the phase diagram of a
substance, as a function of the pressure $P$ and temperature $T$,
showing the generic solid (sol), liquid (liq), and gas phases.  
In order to obtain a liquid wetting film of this substance, the temperature
$T_0$ and pressure $P$ have to be chosen in such a way that the bulk system is
in the vapour phase (fig.~\ref{fig:wetfilm}(b)) but close to the  
condensation transition, occurring at $P = P_0(T_0)$. 
%
%
%
%
%
If the vapour is taken in contact with a suitably chosen (solid) substrate
(hatched grey area in fig.~\ref{fig:wetfilm}(c)), also at temperature $T_0$,
a fluid film will condense on it as a consequence of the interaction of the
substance with the substrate which makes the formation of a liquid
layer thermodynamically favourable. The thickness $L$ of the 
resulting film can be
controlled acting on the undersaturation $\delta P = P - P_0(T_0)$ and 
is determined by van-der-Waals and dispersion forces. 
Depending on the choice of the
substrate, $L$ diverges smoothly 
upon approaching the bulk condensation pressure, i.e.,
for $\delta P \rightarrow 0$. 
This case is referred to as {\it complete wetting} and is
characterized by the formation of a liquid film which can be made
macroscopically thick, so that $L\gg \ell_{\rm micr}$. 
The phenomenon of wetting leads naturally to the formation of a liquid film of
constant thickness $L$, in which the liquid is confined between the solid
substrate and the liquid-vapour interface. The latter is macroscopically well
defined sufficiently far from the liquid-vapour critical point. Even though the
problem of the alignment between the confining surfaces is seemingly solved
by wetting films, one has to be aware of the fact that the liquid-vapour
interface actually fluctuates ({\it capillary fluctuations}) around its
average position. 

A possible {\it indirect} evidence of critical Casimir forces can be obtained
by monitoring the thickness $L$ of the wetting layer upon approaching a
critical point which, in the bulk, occurs within the liquid phase, is close to
the liquid-vapour first-order transition but 
is far enough from the liquid-vapour
critical point. This is the case for {\it critical endpoints} (cep), which are
located where the liquid-vapour transition line terminates a line of critical
points (in the liquid phase, see, c.f., fig.~\ref{fig:He4PD}(a)). 
Indeed, upon approaching a critical endpoint, the critical fluctuations of the
associated order parameter, confined within the liquid wetting film of
thickness $L$, give rise to the critical Casimir effect. 
The associated force adds up to the previously acting (van-der-Waals) forces in
determining the equilibrium distance $L$ of the liquid-vapour interface
from the substrate and consequently it affects the thickness of the wetting
film. 
The dependence of $L$ on the
thermodynamic control parameters such as pressure and temperature is the basis
for the determination of the Casimir force within this approach, which was
proposed and discussed in detail in ref.~\cite{krech:92wet}, providing the
theoretical motivation for the experimental investigations summarized in 
subsec.~\ref{subsec:wf}.

\section{The critical Casimir effect at work}
\label{sec:evid}

In this section we summarize the theoretical predictions for the scaling
function of the critical Casimir force in thin films and we compare them with
the available experimental results based on wetting films
(subsec.~\ref{subsec:wf}). Then we discuss a more direct measurement of
this force (subsec.~\ref{subsec:tirm}).

\subsection{Wetting films}
\label{subsec:wf}
\begin{figure}
\begin{center}
\begin{tabular}{cc}
\begin{minipage}{0.42\textwidth}
\hspace{-6mm}\includegraphics[scale=0.42]{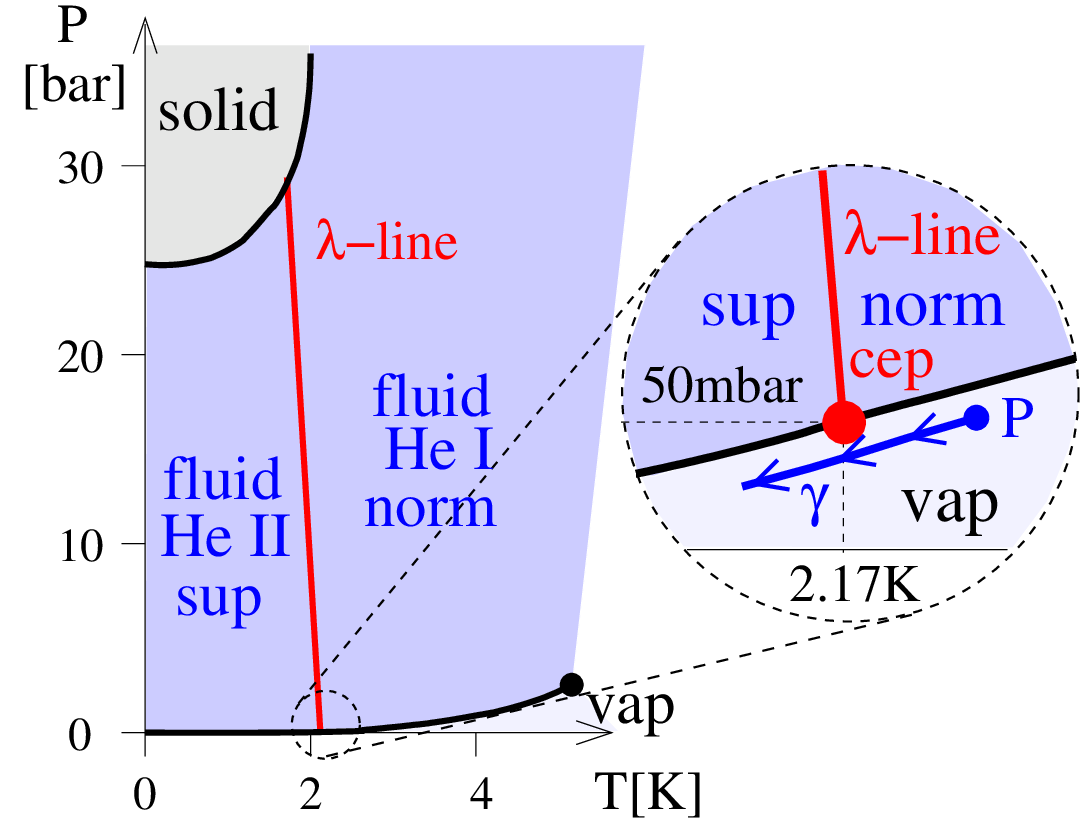} 
\end{minipage}
&
\begin{minipage}{0.58\textwidth}
\includegraphics[scale=1.55]{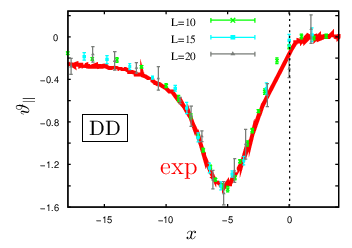} 
\end{minipage}
\\
(a) & (b)
\end{tabular}
\end{center}
\caption{\label{fig:He4PD}%
(a) Phase diagram of pure $^4$He. The closer view highlights the presence of
  the critical endpoint (cep).
(b) Universal scaling function $\vartheta_\|$ of the critical Casimir force
  within the three-dimensional XY universality class and Dirichlet-Dirichlet
  (DD) BCs,  as a function of the scaling variable
  $x=(T/T_\lambda - 1)(L/\xi_0)^{1/\nu}$. The data points were obtained by
  Monte Carlo simulation of the XY model on the lattice in
  ref.~\cite{VGMD-07}, whereas the solid line represents the experimental
  results of ref.~\cite{He4-exp}.
}
\end{figure}

In subsec.~\ref{subsec:cep} we argued that it is possible to exploit critical
endpoints in order to infer the critical Casimir force from the thickness of a
complete wetting film. 
One of the most studied fluids with such a critical
endpoint is pure $^4$He, the phase diagram of which is sketched in
fig.~\ref{fig:He4PD}(a). Within the fluid phase $^4$He undergoes a second-order
phase transition between a {\it sup}erfluid 
(also refereed to as HeII) and a {\it norm}al
(HeI) behaviour upon increasing the temperature across the so-called
$\lambda$-line. The order
parameter $\op$ of this phase transition is physically provided by the
wave-function of the superfluid.  The $\lambda$-line terminates at the critical
endpoint (cep) on the fluid-vapour first-order transition line (which ends on
the right at the liquid-vapour critical point) located at low pressure, as
indicated by the closer view of fig.~\ref{fig:He4PD}(a).
%
%
%
As previously described, in order to detect critical Casimir forces one
prepares the system in the vapour phase corresponding to the
point P in the closer view of fig.~\ref{fig:He4PD}(a) 
and exposes this vapour to
a suitable substrate (e.g., Cu) in order to obtain a complete wetting film of
thickness $L$. Then one changes the thermodynamic parameters in such a way to
follow the thermodynamic path indicated as $\gamma$ in fig.~\ref{fig:He4PD}(a),
along which the film thickness $L$ would not change in the absence of the
critical endpoint. The actual changes are therefore due to the action of the
critical Casimir force on the liquid-vapour interface and from these 
changes it is
possible to determine experimentally the scaling function $\vartheta_\|$ of
the force (see eq.~\reff{eq:critfcas}) \cite{krech:92wet}.
In order to predict theoretically the scaling function $\vartheta_\|$ one has
to single out a theoretical model which belongs to the same {\it bulk} and
{\it surface} universality class as the confined $^4$He.
It is well known that the universal aspects of the bulk critical behaviour
of $^4$He are properly captured, upon approaching the $\lambda$-line, by the 
lattice XY model~\cite{PV}, which lends itself, e.g., for Monte Carlo
simulations. 
Accordingly, the behaviour of a film of $^4$He close to the
$\lambda$-line is captured by the XY model with film geometry, where the
boundary conditions on the lattice are chosen to be in the same surface
universality class (i.e., to lead to the same effective boundary conditions
for the order parameter) as the actual confining boundaries, represented by the
substrate-liquid and liquid-vapour interfaces. 
As previously mentioned, the order parameter $\op$ 
of the superfluid transition is
related to the  wave-function of the superfluid component 
and therefore it has to be continuous in space and to vanish both inside the
substrate and within 
the vapour. As a result, it has to vanish at the boundaries,
i.e., $\op$ satisfies Dirichlet BCs at both
surfaces (DD). On the lattice model these are realized via free boundary
conditions for the degrees of freedom. Having identified a model which belongs
to the same bulk and surface universality class as the film of $^4$He, one can
proceed to the determination of the scaling function $\vartheta_\|$. 
In fig.~\ref{fig:He4PD}(b) the data points refer to the scaling function
$\vartheta_\|$ as inferred from Monte Carlo simulations of the XY model in
film geometry of various thicknesses and DD boundary
conditions~\cite{VGMD-07} (see also ref.~\cite{H-07}). 
The scaling
function is plotted as a function of the proper scaling variable
\be
x= t (L/\xi_0)^{1/\nu}
\label{eq:defx}
\ee
where $t \equiv (T - T_\lambda)/T_\lambda$, 
$T$ is the temperature at which the lattice model is simulated and 
$T_\lambda$ is the associated bulk critical value
(corresponding to the $\lambda$-line in the actual
system). $L$ is the thickness of the film, $\xi_0$ the non-universal
amplitude which controls the  divergence of the bulk correlation length $\xi$
and which can be determined independently via Monte Carlo simulation, and $\nu
\simeq 0.66$~\cite{PV} 
is the universal critical exponent of the correlation length. 
The scaling variable $x$ in eq.~\reff{eq:defx} is related to
$u \equiv L/\xi$ in eq.~\reff{eq:critfcas} by $x = u^{1/\nu}$ for $t>0$. 
As a confirmation of the scaling behaviour in eq.~\reff{eq:critfcas}, the
scaling function $\vartheta_\|$ in fig.~\ref{fig:He4PD}(b) 
does not depend on the
thickness $L$ 
of the film in which it has been numerically determined (as long as
$L$ is large enough compared to the lattice spacing) and the data
sets corresponding to different sizes actually fall onto the same master
curve. 
The predicted critical Casimir force turns out to be attractive in the whole
range of temperatures ($\vartheta_\| < 0$), as expected it decays to zero for
$L \gg \xi$, i.e., for $x \gg 1$, whereas it saturates to a non-vanishing
value for $x \rightarrow - \infty$ due to the fact that at low temperatures
long-range correlations are maintained by Goldstone modes. Some of the
qualitative features of the present force, such as the occurrence of a deep
minimum below $T_\lambda$, can be understood
within a simple mean-field approximation (see, e.g., ref.~\cite{LGW-MF,LGW-MFII}).
%

The very delicate  experimental determination of the critical Casimir force
acting on wetting films of $^4$He on a Cu-substrate has been done in
ref.~\cite{He4-exp}, where the thickness $L\simeq 200\ldots300\,$\AA\ of the
film as a function of the 
distance from the critical point has been determined by capacitance
measurements. The corresponding experimental data are reported as a solid line
in fig.~\ref{fig:He4PD}(b) 
and show a remarkably good agreement with the
corresponding theoretical predictions based on Monte Carlo simulations of the
lattice model. In determining the value of
the abscissa $x$ corresponding to a certain experimental point one has to use
for $\xi_0$ the system-specific value  which has been determined independently
in the experiment on the basis of the behaviour of the bulk 
correlation length. 
If this is properly done, then there are no free parameters which can be
adjusted in the comparison. In this sense the test of the theoretical
predictions is really stringent. 

On the basis of the values of the experimental parameters of
ref.~\cite{He4-exp} one can estimate the critical Casimir
pressure to be of the order of 2\,Pa which, however, is still of quite
difficult detection in those experimental conditions. 
In order to increase the magnitude of the critical Casimir force, which is
proportional to the temperature $T$, it is convenient to look for critical
endpoints occurring at higher temperatures compared to the one of $^4$He,
for example by considering wetting films of 
{\it classical} binary liquid mixture. 
It turns out that in these cases the 
boundary conditions can be tuned in such a way to result in
{\it repulsive} forces. 
These forces could find application for compensating the
attractive  electrodynamic Casimir force (the sign of which is
quite difficult to be controlled) and therefore avoid stiction in micro- and
nano-machines (MEMS and NEMS).

\begin{figure}
\begin{center}
\includegraphics[scale=1.6]{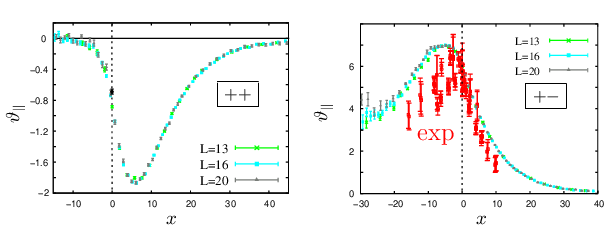}
\end{center}
\caption{\label{fig:Is}%
Universal scaling function $\vartheta_\|$ of the critical Casimir force within
the three-dimensional 
Ising universality class and $(++)/(+-)$ BCs, 
as a function of the scaling variable $x=t(L/\xi_0)^{1/\nu}$. 
The data points were
obtained in ref.~\cite{VGMD-07} 
by Monte Carlo simulation of the Ising model on the lattice, 
whereas experimental data for the case of $(+-)$ BCs were
obtained in ref.~\cite{pershan}.
}
\end{figure}

The phase diagram of a binary liquid mixture is depicted in fig.~\ref{fig:PD}
for a fixed value of the pressure $P$ (i.e., fig.~\ref{fig:PD} is a cut of the
phase diagram in the $(c_A,T,P)$-space). Upon decreasing the pressure the
location of the critical point CP changes in the $(c_A,T)$-plane and
eventually it meets at the critical endpoint the sheet corresponding to the
transition between the liquid and the vapour of the mixture. This critical
endpoint can be exploited in order to create a wetting film of the binary
mixture~\cite{krech:92wet}, 
similarly to the case of $^4$He. The
scaling function $\vartheta_\|$ can therefore be inferred by monitoring the
equilibrium thickness of the wetting layer as a function of the thermodynamic
parameters. 
Also in this case, theoretical predictions for the scaling function
$\vartheta_\|$ can be obtained by studying a suitable 
model which belongs to the same {\it bulk} and {\it surface} universality
class as the confined classical binary mixture. It is  well-known that the
bulk critical behaviour close to the demixing critical point in a binary
mixture is properly captured by the lattice Ising model~\cite{PV}, 
which lends itself,
e.g., for numerical studies via Monte Carlo simulations. The universal
behaviour of the binary mixture 
confined in the film is therefore captured by the Ising
model in a film geometry, in which the boundaries are chosen to be in
the same surface universality class as the actual confining surfaces
constituted by the substrate-liquid and liquid-vapour interfaces. 
In subsec.~\ref{subsec:ccf} we have already mentioned the fact that the
boundaries generically show preferential adsorption for one of the two
components of
the mixture. As a result, the order parameter $\delta c_A({\bf x})$ 
of the binary
mixture (see the discussion in subsec.~\ref{subsec:ccf}) either increases
$(+)$ or decreases $(-)$ upon approaching the boundaries and this tendency
turns into effective boundary conditions close enough to the critical
point~\cite{bind:83,diehl:86:0}. We shall refer to the case in which both
boundaries preferentially adsorb the {\it same} component of the mixture as
$(++)$ and to the case of {\it opposite} preferences as $(+-)$. Additional
details of the strength of this preferences etc. turn out to be irrelevant
sufficiently close to the critical point.
In the lattice Ising model, these boundary conditions are realized by fixing
the boundary degrees of freedom ($\pm 1$ spins) either to the same 
[$(++)$] or to opposite [$(+-)$] values on the two opposing boundaries. 
The resulting numerical prediction of the scaling function $\vartheta_\|$
is reported in fig.~\ref{fig:Is} as a function of the scaling variable $x$
in eq.~\reff{eq:defx}, where $L$ is the thickness of the film, 
$\nu \simeq 0.63$~\cite{PV} 
is the universal critical exponent of the correlation
length, $t = (T-T_c)/T_c$ where $T$ is the temperature at which the lattice
model is simulated and $T_c$ its critical value. As it was the case for
$^4$He, the value of the non-universal amplitude $\xi_0$  for this lattice
model is measured independently via Monte Carlo simulations. 
%
%
%
%
The collapse of data points referring to lattices of different thicknesses
onto the same master curve is again a confirmation of the scaling behaviour
expected on the basis of eq.~\reff{eq:critfcas}\footnote{In order to extract
  in this case the asymptotic scaling behaviour from Monte Carlo data it is
  necessary in to account for corrections to scaling, see ref.~\cite{VGMD-07}
  for details.}.
The sign of the resulting Casimir force depends on the boundary conditions and
is {\it attractive} for $(++)$ and {\it repulsive} for $(+-)$, corresponding,
respectively, 
to $\vartheta_\|(x) < 0$ and  $\vartheta_\|(x) > 0$ in the whole range of
temperatures. As expected, these scaling functions decay to zero for
$x\rightarrow \pm\infty$ due to the fact that away from the critical point the
correlation length becomes much smaller than the film thickness 
$L$ both in the mixed (disordered in terms of the Ising model) and demixed
(ordered) phase.
In addition, the typical 
magnitude of the repulsive force for $(+-)$ BCs
is larger than the one of the attractive force for $(++)$ boundary
conditions. This is due to the fact that in the former case the fluctuations
of the position of the interface between the region with positive and negative
values of $\langle \delta c_A({\bf x})\rangle$ also contribute to
the force acting on the confining surfaces. Also in this case some of the
qualitative features of the force are already captured properly within a
mean-field approximation~\cite{krech:97} 

The critical Casimir effect in wetting films of classical binary mixtures has
been investigated in refs.~\cite{pershan,rafai} where
either ellipsometry~\cite{rafai} or X-ray scattering~\cite{pershan} techniques
have been employed in order to determine the equilibrium thickness
$L$ of the wetting films. In particular in
ref.~\cite{pershan} a binary mixture of methylcyclohexane (MC) and
perfluoromethylcyclohexane (PFMC) with an upper critical point at 
temperature $T_c = 42.6^\circ$C and at a critical molar fraction of PFMC 
$c_{{\rm PFMC},c} = 0.36$ has been
used as the critical medium. The investigated wetting films were formed 
on a SiO$_2$/Si substrate and they had a typical thickness 
$L\simeq 100\,$\AA\ with
the PFMC and MC preferring to be close to the liquid-vapour $(+)$ and the
liquid-substrate  $(-)$ interface, respectively. 
The corresponding experimental data for the inferred scaling function of the
critical Casimir force are reported on the right panel of
fig.~\ref{fig:Is}. 
As in the comparison between theoretical and experimental
predictions for $^4$He, the value of the abscissa $x$
corresponding to a certain experimental point has to be calculated according
to eq.~\reff{eq:defx} taking into
account the experimentally determined value of $\xi_0$ and the fact that $t
\equiv (T - T_c)/T_c$ for an upper critical point. Note that, again, 
there are no adjustable parameters for the comparison. 
The agreement with the corresponding theoretical
predictions is rather good (for a more detailed discussion, see
ref.~\cite{VGMD-07}).

All the evidence previously discussed for critical Casimir forces 
relies on the use of wetting films of (classical and quantum) fluids, 
a rather indirect way of detecting the effects of these forces. 
In addition, the reliable quantitative 
interpretation of the observed variations of the thickness of these 
films in terms of the action of Casimir forces requires an independent and
quite detailed knowledge of some material-dependent parameters which determine
the van-der-Waals forces acting within the system.
%
%
In view of these potential 
limitations, it would be desirable to measure {\it directly} the critical
Casimir forces, e.g., by determining the associated potentials.

\subsection{Brownian motion of colloids}
\label{subsec:tirm}

The experimental evidence summarized in the previous
subsection suggests that critical
Casimir forces become relevant at the submicrometer scale, so that, on
dimensional ground and for critical points at room temperature $T_c\simeq
300\,$K one estimates the typical scale of force as 
$\simeq \kb T_c/(0.1\,\mu{\rm m}) \simeq
40\, $fN. These tiny forces are hardly
detectable even with very sensitive methods such as atomic force microscopy
but nonetheless they are still strong enough to affect the Brownian motion of
particles of micrometer size (i.e., {\it colloids}) which can be used as
detectors. 
This suggests the study of a geometrical arrangement in which the critical
fluctuations of the medium are restricted in the space between a spherical
colloid  and a planar surface rather than between two parallel surfaces,
as considered so far. On the basis of the discussion of
subsec.~\ref{subsec:ccf}, one actually 
expects a Casimir force $F$ to act on the sphere.
If the minimal distance $z$ between the surface of the sphere and
the planar substrate is much smaller than the radius $R$ of the sphere itself,
the spherical surface can be approximated by a set of circular rings
parallel to the opposing planar substrate (Derjaguin approximation) and
therefore the force in
this geometrical setting can be related to the one for parallel plates. Within
this approximation, the $z$-dependent 
critical Casimir force $F$ at temperature $T$ acting 
on the sphere is given by (see ref.~\cite{exp-long} for details)
\be
F(z) 
= \kb T \frac{R}{z^2} \vartheta_{|\circ}(z/\xi)\quad\mbox{for}\quad z \ll R,
\label{eq:fcascoll}
\ee
where the universal scaling function $\vartheta_{|\circ}$ can be calculated in
terms of a suitable integral of the scaling function  $\vartheta_{\|}$ of a 
film~\cite{exp-short} and it retains the qualitative features of
$\vartheta_{\|}$. The theoretical prediction for the scaling
form of the Casimir potential $\Phi_C(z) = \int_z^\infty \rmd z' F(z') = \kb T
(R/z) \Xi(z/\xi) $ immediately follows from the integration of
eq.~\reff{eq:fcascoll}. 

In order to measure tiny forces acting on a single colloid, one can 
take advantage of the Total Internal Reflection Microscopy
(TIRM)~\cite{rev-TIRM}, a very sensitive technique that is widely employed in
the study of soft-matter systems~\cite{rev-TIRM}. With TIRM one monitors the
Brownian motion of the colloid floating in a fluid 
and from the associated statistics one 
infers the potential of the forces acting on the particle. 
%
%
\begin{figure}
\begin{center}
\includegraphics[scale=1.3]{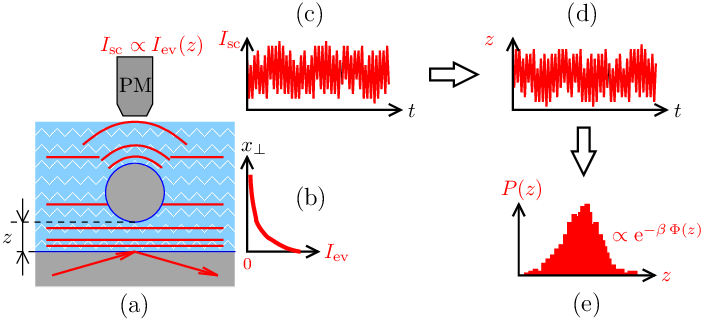}
\end{center}
\caption{\label{fig:TIRM}%
Determination of the potential $\Phi(z)$ of the forces acting on a single 
colloidal particle by means of the Total Internal Reflection Microscopy.
}
\end{figure}
%
%
The scheme of the experimental setup is depicted in fig.~\ref{fig:TIRM}(a): an
incoming visible laser beam is totally reflected at the substrate (glass)-fluid
interface, so that an evanescent field of intensity $I_{\rm ev}(x_\bot)$
penetrates the liquid with a typical exponential decay as a function of the
distance $x_\bot$ from the interface, see fig.~\ref{fig:TIRM}(b).   
The colloidal particle, of suitable refractive index and 
at a surface-to-surface distance $z$ from the substrate, scatters light out of
this evanescent field. The intensity $I_{\rm sc}$ of the scattered light is
measured by a photomultiplier (PM) and turns out to be proportional to the
intensity  $I_{\rm ev}(z)$ of the evanescent field.
From the time dependence of the scattered intensity
$I_{\rm sc}(t) \propto I_{\rm ev}(z(t))$, represented in
fig.~\ref{fig:TIRM}(c), one can therefore calculate the time dependence
of the particle-wall separation, i.e., $z(t)$.
This dependence, depicted in  fig.~\ref{fig:TIRM}(d), reflects the
Brownian motion of the particle under the effects of thermal fluctuations and
possible additional forces. 
The probability distribution $P(z)$, as extracted
from the histogram of $\{z(t)\}_{0\le t \le t_{\rm samp}}$ for sufficiently
large sampling time $t_{\rm samp}$ (fig.~\ref{fig:TIRM}(e)), is
proportional  to the Boltzmann factor $\exp\{-\beta\Phi(z)\}$ (with $\beta =
(\kb T)^{-1}$), where $\Phi(z)$ is the potential of the total force acting on
the particle. Accordingly, $\Phi(z)$ can be calculated up to an irrelevant
constant on the basis of the $P(z)$ inferred from the experimental data. 

In order to provide a {\it direct} evidence of critical Casimir forces one can
study a single colloidal particle immersed in a binary liquid mixture of
water (W) and lutidine (L), close to a glass substrate~\cite{exp-short}. 
In ref.~\cite{exp-short} TIRM has been employed for the determination of the
potential  $\Phi(z)$ of the forces acting on the particle and for studying how 
$\Phi$ changes when the binary mixture is driven towards the critical point.
The W-L mixture has a bulk phase
diagram such as the one presented in fig.~\ref{fig:PD}, with a lower critical
point at $T_c = 34^\circ$C and lutidine mass fraction $c_{{\rm L},c} \simeq
0.29$. The local excess concentration of
lutidine $\delta c_{{\rm L}}({\bf x})$ is a suitable order parameter for this
demixing transition. 
In what follows we illustrate the results of the experiments in the case 
of a mixture at the critical concentration $c_{{\rm L},c}$ (the case $c_{{\rm
    L}} \neq c_{{\rm L},c}$ is discussed in refs.~\cite{exp-long,exp-short}).

\begin{figure}[h!!]
\begin{center}
\begin{tabular}{cc}
\hspace{-6mm}\includegraphics[scale=0.85]{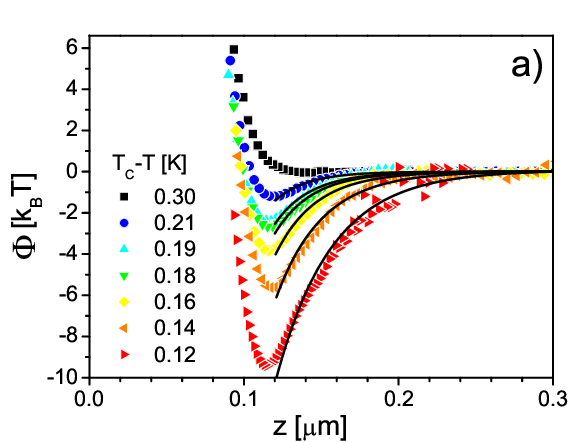} &
\includegraphics[scale=0.85]{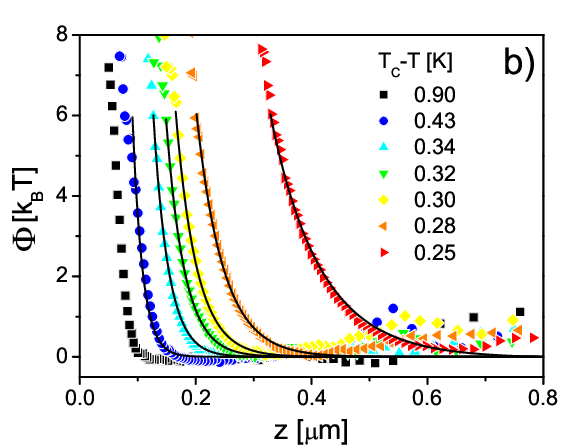}
\end{tabular}
\end{center}
\caption{\label{fig:pots}%
Temperature dependence of the potential $\Phi(z)$ of the forces acting on
single polystyrene colloidal particles at a distance $z$ of closest approach 
from a glass substrate. The particles and the substrate are immersed in a
water-lutidine liquid mixture with critical composition and temperature $T$
smaller than the critical one $T_c$. Data points refer to the experimental
results (where gravity and buoyancy have been subtracted, see the main text)
obtained for a substrate treated with NaOH in order to make it hydrophilic
[$(-)$ BC for the order parameter] and
(a) a colloid of diameter $2R = 2.4\,\mu$m absorbing preferentially water
$(-)$ (b) a colloid of diameter $2R = 3.7\,\mu$m with preferential
adsorption for lutidine $(+)$. The solid lines in (a) and (b) are the
corresponding theoretical predictions for the critical Casimir potential with
$({\rm colloid},{\rm substrate}) = (--)$ and $(+-)$ BCs,
respectively. (From ref.~\cite{exp-short}.)
}
\end{figure}

At a temperature $T$ far below the critical value $T_c$, 
fluctuations of the order parameter
are irrelevant and the potential $\Phi(z)$ of the forces acting on the
particle is basically the sum of two
contributions: (a) a screened electrostatic repulsion ($\Phi_{\rm
  el}(z)\propto \rme^{-\kappa z}$)
between the colloid and the glass
substrate, relevant in our experimental conditions for $z\lesssim 0.1\,\mu$m 
and (b) a linearly increasing potential due to the combined effect of buoyancy,
gravity and optical pressures generated by the optical tweezer employed in the
actual experimental setup\footnote{A third possible contribution, due to
  van-der-Waals forces turned out to be negligible in the experiment reported
  in ref.~\cite{exp-short,exp-long}.}. %
This latter contribution has been subtracted from
the curves presented in fig.~\ref{fig:pots}, which show the dependence 
of the measured potential $\Phi(z)$ on the distance from the
critical point. 
%
%
%
%
%
Panel (a) and (b) in fig.~\ref{fig:pots} 
refer to $({\rm colloid},{\rm substrate}) = (--)$ and $(+-)$
BCs, respectively. As expected, few hundreds mK far from the
experimentally estimated critical temperature, the potential $\Phi(z)$
consists only of the electrostatic repulsion, which is negligible on this scale
for $z\gtrsim 0.12\,\mu$m in (a) and $z\gtrsim 0.1\,\mu$m in (b). Upon
increasing the temperature by few tens mK significant changes occur in the
measured potentials as a strong attractive and repulsive force develops,
respectively, in (a) and (b). In particular, in both cases one observes 
a change
in the total potential of the order of 10 $\kb T$, which is really significant
on the scale of $\kb T$ which characterizes the physics of colloids. The
maximum value of the attractive force acting in case (a) is estimated at 600
fN. The strong temperature dependence of the measured potentials is a clear
indication of the involvement of critical Casimir forces. This evidence is
supported by the comparison with the corresponding theoretical predictions for
$\Phi_C(z)$, shown as solid lines in fig.~\ref{fig:pots}. 
In order to focus only on the contribution to $\Phi(z)$ due to the critical
Casimir force, this comparison is presented only within the range 
of distances where the electrostatic contribution is negligible.
The limited accuracy with which  the critical temperature $T_c$
has been experimentally determined in ref.~\cite{exp-short}
does not allow a reliable calculation of the correlation length $\xi$ 
via $\xi = \xi_0(1-T/T_c)^{-0.63}$ and on the
basis of the value $\xi_0^{\rm (exp)}$ 
of $\xi_0$ measured for the W-L mixture in the bulk 
by independent experiments~\cite{xi0}.
Instead, for each single temperature $T$, 
the correlation length $\xi(T)$ can be determined in order to 
optimize the agreement between theory and experimental data. 
Then, separately for the two experiments,
the value  $\xi_0^{\rm (fit)}$ of $\xi_0$ and $T_c^{\rm (fit)}$ of $T_c$ are
determined in such a way to
yield the best fit of the determined $\xi(T)$ with the expected algebraic
behaviour. In both cases the fact that, within errorbars, 
$\xi_0^{\rm (fit)} = \xi_0^{\rm (exp)}$
is a check of the significant agreement between theory and experiments. (See
refs.~\cite{exp-long,exp-short} for a more detailed discussion.)
Interestingly enough, via a suitable chemical surface treatment, it is
possible to change quite easily the substrate from hydrophilic $(-)$ to
hydrophobic $(+)$ and therefore switch the Casimir force acting on the
colloids in fig.~\ref{fig:pots}(a) and (b), respectively, 
from attractive to repulsive and
vice-versa (see refs.~\cite{exp-long,exp-short}).

Summing up, by measuring their effect on the Brownian motion of a colloidal
particle, direct evidence of both attractive and repulsive critical
Casimir forces has been provided. Based on the general arguments presented in
subsec.~\ref{subsec:ccf} we expect such forces to act also between two or more
colloids immersed in a near-critical mixture. Due to the strong non-additivity
of these fluctuation-induced forces we expect interesting many-body
effects. 

\section{Conclusions, perspectives and applications}
\label{sec:persp}

\subsection{Conclusions}
In the previous sections we presented an overview of the Casimir effect due to
critical fluctuations, highlighting its relation with the analogous effect in
QED. We focussed on the theoretical predictions which are relevant
for the interpretation of the available and possibly forthcoming experiments,
especially involving wetting films and colloidal particles. 
The comparison
between the available theoretical and experimental results turns out to be
very good. 
We did not attempt, however, to review of all the currently available
relevant theoretical results and approaches, a task which is well beyond the
scope of the present contribution. In what follows we mention only some of
them. 

The discussion in
subsec.~\ref{subsec:ccf} clearly shows that Casimir-like effects are expected
whenever the spectrum of fluctuations of a certain nature is changed by the
presence of confining bodies, which consequently experience
fluctuation-induced forces. 
The range of such forces is typically set by the correlation length $\xi$ of
the relevant fluctuations and a {\it universal} behaviour has generally to be
expected whenever $\xi, L \gg \ell_{\rm micr}$ where $L$ is the typical
distance between the boundaries and $\ell_{\rm micr}$ a microscopic,
system-dependent scale. 

\subsection{Perspectives}

In view of the quite general conditions under which such Casimir forces
arise, several examples can be found in statistical physics. 
They occur not only at the critical points previously discussed, or in
systems belonging to other bulk and surface universality classes (e.g., liquid
crystals or complex fluids~\cite{KG-99,LC-CF}), 
but also close to {\it tri}critical
points. This instance has been studied both experimentally and
theoretically~\cite{He-mix,LGW-MF}. 
Due to the fact that the upper critical spatial dimensionaliy for a
tricritcal point is three, a mean-field approximation of suitable models
captures rather well the features observed in experiments. Casimir-like forces
result also from (long-ranged) fluctuations occurring in non-equilibrium
steady states of different nature, from chemical reactions to granular matter,
see, e.g., ref.~\cite{Neq-Casimir}, or from quantum fluctuations of the order
parameter of a quantum phase transition~\cite{D-book}. However, at present it
is not clear how these predictions can be experimentally tested. 
On the other hand, there are experimentally relevant systems, such as Bose
gases close to the Bose-Einstein condensation, which have not yet been studied
with the aim of detecting critical Casimir forces, even though some
quantitative theoretical predictions (see fig.~\ref{fig:He4PD}(b)) 
are available beyond the case of ideal gases~\cite{EPL-comm}.
Thermal capillary fluctuations of the interface between two different liquids
give rise to a Casimir force acting on two or more collodal particles trapped
at that interface~\cite{capillary}. In this case, the presence of the colloids
alters the spectrum of these surface fluctuations, with the additional feature 
that the boundaries themselves at which the boundary conditions are imposed,
represented by the contact line of the interface at the surface of the
colloids, fluctuate (see ref.~\cite{capillary} for details). The resulting
forces might be of relevance for understanding 
the topical problem of the interaction among
colloids trapped at interfaces. 

%
%
%
\begin{figure}[h!!]
\begin{center}
\includegraphics[scale=1.2]{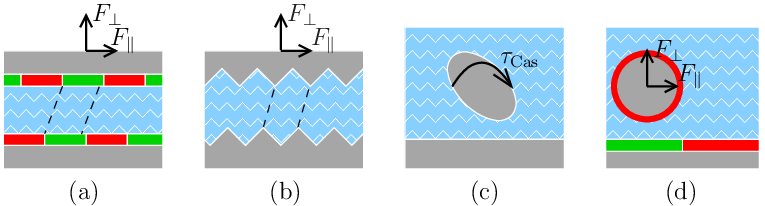}
\end{center}
\caption{\label{fig:patt}%
Possible chemically [(a), (d)] and topographically [(b)] patterned substrates,
in which a lateral critical Casimir force $F_\|$ arises in addition to the
normal force $F_\bot$ previously discussed. A  critical Casimir torque
$\tau_{\rm Cas}$ is expected to act on non-spherical particles [(c)].
}
\end{figure}
%
%

So far we have only considered homogeneous substrates and particles.
However, in the presence of 
chemical or topographical modulations, such as the one
depicted in fig.~\ref{fig:patt}, novel phenomena
are expected, including the emergence of lateral critical Casimir forces $F_\|$
and torques $\tau_{\rm Cas}$, still characterized by a universal scaling
behaviour. Some of the features of these forces have been studied in
refs.~\cite{non-hom} and might form the basis for the quantitative
understanding of recent experimental investigation of 
the behaviour of a single colloidal particle immersed in a binary mixture and
exposed to a patterned substrate~\cite{non-hom-exp}.

In addition to the equilibrium static behaviour of the critical Casimir
effect, its equilibrium and non-equilibrium dynamics poses both a theoretical
and experimental challenge.
Indeed, it is not even obvious how to define theoretically such forces when
their definition based  on the decomposition of the effective free energy
$\FF$ (see eq.~\reff{eq:fsp}) is no longer viable due to the lack of a proper
notion of free energy for dynamical phenomena. 
A rich dynamical behaviour is actually emerging already
in quite simple theoretical relaxational models~\cite{dyn}. In order to make
contact with the dynamical information which might come from TIRM experiments,
it remains to be seen how the time-dependence of the critical Casimir force 
affects the Brownian motion of the colloidal particle in conjunction with
hydrodynamic and adsorption phenomena.

\subsection{Applications?}

In contrast with the interactions typically acting among colloids (e.g.,
elctrostatic), the critical Casimir forces show a striking temperature
dependence, as we have reported in subsec.~\ref{subsec:tirm}. This fact
can possibly be exploited in order to control via minute temperature changes
the phase behaviour and aggregation phenomena in systems with dispersed
colloids. 
In addition, not only the range of interaction can be controlled but also the
sign of the resulting force. Typically this can be achieved by surface
treatments and does not require (as it does for the quantum mechanical Casimir
effect) substantial changes or tuning of the properties of the bulk materials
which constitutes the immersed objects. 
This property might be exploited in order to neutralize the attractive quantum
mechanical Casimir force responsible for the stiction which brings
micro-electromecanical systems to a standstill. If these machines would work
not in a vacuum but in a liquid mixture close to the critical point,
the stiction could be prevented by tuning the critical Casimir force to be
repulsive via a suitable coating of the various machine parts. 
In principle, by using optically removable or controllable coatings, one could
very conviniently control the functioning of the microdevice without acting
directly on it.

\ack
I am grateful to 
C. Bechinger, S. Dietrich, L. Helden, C. Hertlein, A. Macio\l ek and
O.~Vasilyev for the stimualting collaborations which lead to some of the
results summarized here and to M.~Oettel and D.~Dantchev 
for a careful reading of the manuscript.


\newcommand{\rev}[4]{{#3} {\it #1} {\bf #2} #4}

\end{document}